\documentclass[a4paper,12pt]{article}
\usepackage[utf8x]{inputenc}
\usepackage{a4wide,color}
\usepackage{amsmath}
\usepackage{graphicx,epsf}

\def\beq{\begin{eqnarray}}
\def\eeq{\end{eqnarray}}

\begin{document}

\begin{titlepage}
\begin{flushright}
DO-TH 15/11\\
QFET-2015-27\\
 arXiv:YYMM.NNNNvV\\
 \today
\end{flushright}
\vskip 2.4cm

\begin{center}
\boldmath
{\Large\bf Higher-order Wilson coefficients for c $\to$ u transitions in the Standard Model}
\unboldmath
\vskip 2.2cm
{\sc Stefan de Boer$^a$}, \hspace*{0.3cm}{\sc Bastian M\"uller$^b$}
\hspace*{0.3cm}and\hspace*{0.3cm}{\sc Dirk Seidel$^b$}
\vskip .5cm

\vspace{0.7cm}
{\sl ${}^a$Fakult\"at f\"ur Physik,\\
TU Dortmund, Otto-Hahn-Str.4, 44221 Dortmund, Germany\\
\vspace{0.3cm}
${}^b$Theoretische Physik 1,\\
Universit\"at Siegen, Walter-Flex-Str. 3, 57068 Siegen, Germany}
\vspace{3\baselineskip}

\vspace*{0.5cm}

\end{center}

\begin{abstract}

The standard theoretical framework to deal with weak decays of heavy mesons is the so-called 
weak effective Hamiltonian. It involves the short-distance Wilson coefficients, which depend on the 
renormalisation scale $\mu$. For specific calculations one has to evolve the Wilson coefficients down 
from the electroweak scale $\mu=M_W$ to the typical mass scale of the decay under consideration. This is 
done by solving a renormalisation group equation for the effective operator basis. In this paper the 
results of a consistent two-step running of the $c \to u \,\ell^+\ell^-$ Wilson coefficients for 
dimension-6 operators are presented. This running involves the intermediate scale $\mu=m_b$ (with $M_W > 
m_b > m_c$) where the bottom quark is integrated out. The matching coefficients and anomalous dimensions 
are taken to the required order by generalizing and extending results from $b \to s$ or $s \to d$ 
transitions available in the literature.

\end{abstract}

\end{titlepage}

\section{Introduction}

The study of flavour-changing neutral current (FCNC) transitions is a key tool to explore the
generational structure of standard model (SM) fermions, and to look for physics beyond the standard
model (BSM). A lot of work has been done to analyse processes involving $b$-quarks where in the meantime 
theoretical predictions and experimental measurements have reached a high level of 
precision~\cite{Blake:2015tda}. In contrast to that, investigations of charm FCNCs are much less 
advanced due to several reasons. The corresponding rates are highly 
GIM-suppressed~\cite{Glashow:1970gm}, experimental analyses are challenging, and decay modes are 
subjected to resonance contributions, shielding the electroweak physics. In many cases, extensions of the 
SM may upset the GIM suppression and give contributions which are sometimes orders of magnitude larger 
than within the SM. 

Due to the specific CKM and mass structure of charm FCNCs, also the electroweak contributions within the 
SM can differ by several orders of magnitude depending on which corrections are taken into 
account~\cite{Greub:1996wn}. It is therefore desirable to extend the SM calculation for the $c\to 
u\ell^+\ell^-$ transition to ${\cal O}(\alpha_s)$ within renormalisation-group improved perturbation 
theory. As a first step the weak effective Hamiltonian consisting of all relevant dimension-6 operators 
with the corresponding Wilson coefficients is needed to this order. In this paper we will present results 
for this step at next-to-next-to-leading logarithmic (NNLL) order which is required for a consistent 
treatment of the decays at ${\cal O}(\alpha_s)$.

The calculation of the Wilson coefficients is in many parts analogous to the one in the $B$-meson 
sector. The main difference is that in the case considered here, we have to perform a two-step matching. 
In addition to the matching at the high scale $M_W$, the bottom-threshold is crossed when evolving the 
renormalisation scale down to the charm mass. Therefore the bottom-quark has to be integrated out which 
leads to non-trivial matching conditions at the scale $\mu=m_b$. The running of the coefficients at 
the intermediate steps $M_W>\mu>m_b$ and $m_b>\mu$ can be performed analogously to the decay $b\to 
d/s\,\ell^+\ell^-$, where only the charge assignments and the number of active flavours of the 
corresponding anomalous dimensions have to be adapted accordingly. The matching conditions at the high 
scale and the anomalous dimensions are known at the NNLL 
order~\cite{Buchalla:1995vs,Borzumati:1998tg,Chetyrkin:1996vx,Chetyrkin:1997gb,Bobeth:1999mk,
Gambino:2003zm,Gorbahn:2004my,Gorbahn:2005sa}.

In the next section we will present the effective Hamiltonian relevant for $c\to u$ transitions. The 
matching conditions at the high scale $M_W$ and the relevant formulae for the running down to the charm 
scale are given. As some of the anomalous dimension matrices are only presented with explicit 
assignments for the quark charges and number of flavours for bottom decays in the literature, we will 
present them with the full parameter dependence. At the end of that section, the numerical values of the 
Wilson coefficients at the charm-mass scale are given and will be compared to the corresponding 
coefficients 
for $b$-decays. In section 3 we will focus on the clarification of some misunderstanding present in 
previous work. We will therefore present the effective Wilson coefficient  $C_9^{\rm eff}$ at order 
$\alpha_s^0$ and compare the results with existing treatments in the literature. Finally, in the 
appendix, we give formulae to switch between different operator bases for the effective weak Hamiltonian.

\section{Effective Hamiltonian for $c \to u \ell \ell$}

The short-distance expansion has to be divided into two steps: Firstly, we integrate out the weak gauge 
bosons at a scale $\mu_W \sim M_W$. At this step, there are no penguin operators generated, as all 
$d$-type quark masses should be treated as massless~\cite{Greub:1996wn} and the GIM 
mechanism is in full effect. The effective Hamiltonian for scales 
$M_W > \mu > m_b$ is given by
\begin{eqnarray}
H_{\rm eff}(M_W > \mu > m_b) &=& \frac{4 G_F}{\sqrt{2}} \sum_{q=d,s,b} 
V^*_{cq} V_{uq} [ C_1(\mu) {\cal O}^q_1 + C_2(\mu) {\cal O}^q_2 ]\, , \label{H1}
\end{eqnarray}
where
\begin{eqnarray}
 {\cal O}^q_1 &=& (\bar{u}_L \gamma_{\mu} T^a q_L)
          (\bar{q}_L \gamma^{\mu} T^a c_L)\, ,\\
{\cal O}^q_2 &=& (\bar{u}_L \gamma_{\mu} q_L) 
          (\bar{q}_L  \gamma^{\mu} c_L)\, ,
\end{eqnarray}
$T^a$ are the generators of SU(3), and the subscript $L$ denotes left-handed fields.
Secondly, one integrates out the bottom-quark around $\mu_b \sim m_b$. This generates penguin 
operators with Wilson coefficients depending on $M_W$ solely through $C_{1,2}(m_b)$. 
The effective Hamiltonian for scales $m_b > \mu > m_c$ is thus given by
\begin{eqnarray}
H_{\rm eff}(m_b > \mu > m_c) &=& \frac{4 G_F}{\sqrt{2}} \sum_{q=d,s} 
V^*_{cq} V_{uq} [ C_1(\mu) {\cal O}^q_1 + C_2(\mu) {\cal O}^q_2 
 + \sum_{i=3}^{10} C_i(\mu) {\cal O}_i]\, ,\label{H2}
\end{eqnarray}
with
\begin{eqnarray}
{\cal O}_3 & = &(\bar{u}_L \gamma_\mu c_L) \sum_{q=u,d,s,c} (\bar{q} \gamma^\mu q) \, ,\\
{\cal O}_4 & = &(\bar{u}_L \gamma_\mu T^a c_L) \sum_{q=u,d,s,c} (\bar{q}\gamma^\mu T^a q) \, ,\\
{\cal O}_5 & = &(\bar{u}_L \gamma_\mu \gamma_\nu \gamma_\rho c_L)
\sum_{q=u,d,s,c} (\bar{q} \gamma^\mu \gamma^\nu \gamma^\rho q) \, ,\\
{\cal O}_6 & = &(\bar{u}_L \gamma_\mu \gamma_\nu \gamma_\rho T^a c_L)
\sum_{q=u,d,s,c} (\bar{q} \gamma^\mu \gamma^\nu \gamma^\rho T^a q) \, ,\\
{\cal O}_7 & = &-\frac{g_{\rm em}m_c}{16\pi^2} (\bar{u}_L \sigma^{\mu \nu} c_R) F_{\mu \nu}\, ,\\
{\cal O}_8 & = &-\frac{g_s m_c}{16\pi^2} (\bar{u}_L \sigma^{\mu \nu} T^a c_R) G_{\mu
  \nu}^a \, ,\\
{\cal O}_9 & = &\frac{\alpha_{\rm em}}{4\pi} (\bar{u}_L \gamma_\mu c_L) (\bar{\ell} \gamma^\mu \ell)\,,\\
{\cal O}_{10} & = & \frac{\alpha_{\rm em}}{4\pi} (\bar{u}_L \gamma_\mu c_L)(\bar{\ell} \gamma^\mu 
\gamma_5 \ell)\, .
\end{eqnarray}
The sign convention for ${\cal O}_{7,8}$ corresponds to $+ig_sT^a$, $+ig_{\rm em} e_f$ for the ordinary 
quark-gauge-boson vertex ($e_f = −1$ for charged lepton fields).
Only $C_{1/2}$ receive non-zero contributions from the matching procedure at $\mu\sim M_W$, all 
Wilson coefficients of the penguin operators vanish identically as noted above. As a consequence 
$C_{3-9}$ receive non-zero contributions only from the matching of the five-flavour effective theory 
above the scale $m_b$ to the four-flavour effective theory below that scale and from the  mixing of 
${\cal O}_{1/2}$ into ${\cal O}_{3-9}$ below the scale $m_b$, where the $b$-quark has been integrated 
out. $C_{10}$ does not mix under renormalisation and thus is zero at all scales to leading order in the 
$1/M_W$ expansion.

Our aim is to determine the Wilson coefficients at a perturbative order which is suitable for performing 
analyses for $D$-decays\footnote{The results presented here can of course also 
be used for charmed baryons like $\Lambda_c$.} at first order in the strong coupling $\alpha_s$. Because 
the anomalous dimension of 
${\cal O}_9$ begins 
at order $\alpha_s^0$, the Wilson coefficient $C_9$ is needed to NNLL accuracy. This requires also the 
coefficients of the four-quark operators to this accuracy. At 
the scale $\mu\sim m_c$, the Wilson coefficients $C_{1-8}$ are given by
\begin{equation}\label{eq::RunC8}
 C(\mu) = U^{(n_f=4)}(\mu,m_b)\,R\,U^{(n_f=5)}(m_b,M_W) \,C(M_W)\,,
\end{equation}
where $C(\mu)$ is to be understood as the vector of Wilson coefficients. In the following we will not 
present the results 
for the coefficients $C_{7/8}$, but rather for the renormalisation-scheme independent effective ones 
defined by
\begin{eqnarray}
 C_{7/8}^{\rm eff}(\mu) = C_{7/8}(\mu) + \sum_{i=1}^6 y_i^{(7/8)}C_i(\mu)\,,
\end{eqnarray}
with $y^{(7)}=Q\, (0,0,1,\frac43,20,\frac{80}{3})$ and 
$y^{(8)}=(0,0,1,-\frac16,20,-\frac{10}{3})$ in the chosen operator basis. One has to make the 
assignments $Q=Q_u=2/3$ and $Q=Q_d=-1/3$ for $D$-decays and $B$-decays, respectively.

$U^{(n_f)}(\mu_1,\mu_2)$ is the 
evolution matrix 
which includes the 
renormalisation-group improved contributions from the scale $\mu_2$ down to $\mu_1$ and 
$R$ is the 
matching matrix between the five- and four-flavour effective theory. 
As noted above, the vector containing $C_{1-8}$ at the scale $M_W$, $C(M_W)$, has only two non-zero 
entries, which are given by~\cite{Bobeth:1999mk}
\begin{eqnarray}
 C_1(M_W) &=& 15 a_s + a_s^2\left[(16x+8)\sqrt{4x-1} \; {\rm Cl}_2\left(2 \arcsin 
   \frac{1}{2\sqrt{x}}\right) \right.\\\nonumber
  &&\left.\hspace{5em}-\left(16x+\frac{20}{3}\right) \ln x - 32x+ \frac{7091}{72} +\frac{17}{3} 
\pi^2\right],\\
C_2(M_W) &=& 1 + a_s^2\left(\frac{127}{18} +\frac{4}{3} \pi^2\right),
\end{eqnarray}
where $x=[\hat m_t(M_W)/M_W]^2$ with the top quark $\overline{\rm MS}$-mass $\hat m_t$ and
$a_s=\alpha_s/(4\pi)$. The Clausen-function is defined as
\begin{eqnarray}
{\rm Cl}_2(x) &=& {\rm Im}\left[ {\rm Li}_2(e^{ix}) \right],
\end{eqnarray}
with the dilogarithm ${\rm Li}_2$. The evolution matrix
$U^{(n_f)}(\mu_1,\mu_2)$ satisfies
\begin{equation}
\frac{d}{d\ln\mu_1} \,U^{(n_f)}(\mu_1,\mu_2) = \gamma^T(n_f,\mu_1) \, U^{(n_f)}(\mu_1,\mu_2)\,.
\end{equation}
The solution for this matrix at NNLL order is given in~(C.6) in~\cite{Beneke:2001at} for 
$B$-decays. Trivial changes have to be incorporated for the case considered here. The anomalous 
dimension matrix is expanded as
\begin{equation}
\gamma(n_f,\mu_1) = \gamma^{(0)} a_s(n_f,\mu_1)+\gamma^{(1)} a_s(n_f,\mu_1)^2+\ldots\, .
\end{equation}
The $6\times 6$ 
submatrix of the anomalous dimension with full $n_f$ dependence can be found 
in~\cite{Chetyrkin:1997gb,Gorbahn:2004my}. The $2\times 2$ submatrix from self-mixing in the dipole 
operator sector is given 
in~\cite{Gorbahn:2005sa}. This matrix depends also on the charges of the quarks, which have to be chosen 
appropriately for the case of $D$-decays considered in this paper. Up to the required order, the 
$6\times 2$ submatrix from 
mixing between four fermion and dipole operators has only been given in the 
literature for $B$-decays~\cite{Gambino:2003zm}. With the 
full dependence on the charges and active flavours it reads~\cite{MGprivate}
\begin{equation}
 \gamma^{({\rm eff},0)}_{6\times 2}=\left(
\begin{array}{cc}
 -\frac{4}{3} q_1-\frac{8}{81} q_2 & \frac{173}{162} \\
 8 q_1+\frac{16}{27} q_2& \frac{70}{27} \\
 \frac{176}{27} q_2 & \frac{14}{27} \\
 \left(-\frac{88}{81}+\frac{16}{27} n_f\right) q_2 & \frac{74}{81}-\frac{49}{54} n_f \\
 \frac{6272}{27} q_2 & \frac{1736}{27} + 36 n_f\\
 48 n_1 \bar q+\left(-\frac{3136}{81}+\frac{1456}{27} n_f\right) q_2 & 
   \frac{2372}{81} + \frac{160}{27} n_f\\
\end{array}
\right),
\end{equation}

\begin{eqnarray}
 \gamma^{({\rm eff},1)}_{6\times 2}=\left(
\begin{array}{cc}
 \left(-\frac{374}{27} +\frac{2}{27}n_f\right) q_1+\left(-\frac{12614}{729}
  +\frac{64}{729}n_f\right) q_2 
  & \frac{65867}{5832} + \frac{431}{5832}n_f\\
 \left(\frac{136}{9} -\frac{4}{9}n_f\right)
   q_1+\left(\frac{2332}{243}-\frac{128}{243} n_f\right) q_2
  & \frac{10577}{486}-\frac{917}{972} n_f\\
 -\frac{112}{3} n_1 \bar q -\left(\frac{97876}{243}+\frac{4720}{243} n_f\right) q_2
  & \frac{42524}{243}-\frac{2398}{243} n_f\\
   -\frac{140}{9} n_1 \bar q 
         +\left(\frac{70376}{729}+\frac{4448}{729} n_f-\frac{32}{243} n_f^2\right) q_2
  & -\frac{159718}{729} -\frac{39719}{5832} n_f-\frac{253}{486}n_f^2\\
  -\frac{3136}{3} n_1 \bar q -\left(\frac{1764752}{243}+\frac{188608}{243} n_f\right) q_2
  & \frac{2281576}{243}+\frac{140954}{243}n_f -14 n_f^2\\
   \gamma^{({\rm eff},1)}_{67}
  & -\frac{3031517}{729} -\frac{15431}{1458}n_f-\frac{6031}{486} n_f^2\\
\end{array}
\right),
\end{eqnarray}
with
\begin{eqnarray}
 \gamma^{({\rm eff},1)}_{67} &=& -\left(\frac{1136}{9} + \frac{56}{3}n_f \right) n_1 \bar q 
         -\left(\frac{4193840}{729}-\frac{232112}{729} n_f+\frac{5432}{243} n_f^2\right) q_2\,,
  \nonumber\\
  \bar q &=& q_1 - q_2\,.
\end{eqnarray}
For the case of $D$-meson decays one has to make the assignments $q_1=Q_d=-1/3$, $q_2=Q_u=2/3$, $n_2=2$ 
and $n_1=3$ ($n_f=5$) or  $n_1=2$ ($n_f=4$). The matrices given in the literature are reproduced with 
the following assignment for $B$-decays: $q_1=Q_u=2/3$, $q_2=Q_d=-1/3$, $n_2=3$ and $n_1=2$.

\begin{figure}[t]
\begin{center}
\includegraphics[width=10em, keepaspectratio]{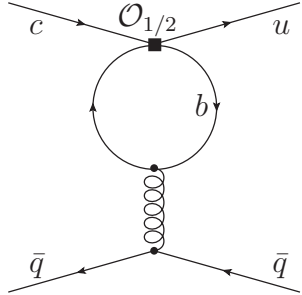}
\put(-105,10){$\bar q$}
\put(-14,10){$\bar q$}
\put(-105,100){$c$}
\put(-14,100){$u$}
\put(-43,69){$b$}
\put(-72,103){${\cal O}_{1/2}$}
\end{center}
\centerline{\parbox{0.8\textwidth}{\caption{\label{diag::match}\noindent Diagrams relevant for the 
matching of the five-quark to the four-quark effective theory at order $\alpha_s$.}}}
\end{figure}

The matrix $R$ in (\ref{eq::RunC8}) is the matching matrix from the five to the four 
active flavour effective theory. It is different from the unit matrix because the operators 
${\cal O}_{1/2}^b$ are absent below the $b$-quark threshold. It is given by
\begin{eqnarray}
 R_{ij} = \delta_{ij} + a_s(m_b)\, R_{ij}^{(1)} + a_s(m_b)^2\, R_{ij}^{(2)} + \dots\,.
\end{eqnarray}
At order $\alpha_s$ the non-zero elements of $R_{ij}^{(1)}$~\cite{Buras:1993dy} are obtained from the 
diagrams depicted in Fig.~\ref{diag::match} at zero momentum transfer\footnote{Note that this is valid 
for the effective Wilson coefficients $C_{7/8}^{\rm eff}$.}:
\begin{eqnarray}
 R_{41}^{(1)}&=&-R_{42}^{(1)}/6=1/9\,,\nonumber\\
  R_{71}^{(1)}&=&-R_{72}^{(1)}/6=8/81\,, \qquad
 R_{81}^{(1)}=-R_{82}^{(1)}/6=-1/54\,.
\end{eqnarray}
The contributions at order $\alpha_s^2$ are not known yet. The diagrams including an additional gluon 
connecting only the upper fermion lines in Fig.~\ref{diag::match} have been calculated for $B$-physics 
in~\cite{Asatryan:2001zw}. Unfortunately the calculation involves an expansion in $m_c/m_b$, which in 
the case considered here would turn into an expansion in $m_b/m_c$ and is thus not applicable. In the 
following we will set $R^{(2)}\simeq 0$ as an approximation.

For $C_9$ we get the following evolution down to the scale $\mu\sim m_c$:
\begin{equation}
C_9(\mu) = C_9(m_b) + W^{(n_f=4)}(\mu,m_b)\,R\,U^{(n_f=5)}(m_b,M_W)\,C(M_W)\,,
\end{equation}
with the $1\times 6$ matrix
\begin{equation}\label{eq::W}
 W^{(n_f=4)}(\mu,m_b) = -\frac{1}{2}\,\int_{a_s(m_b)}^{a_s(\mu)}\!da_s 
\,\frac{\kappa(a_s)}{\beta(a_s)}\,U^{(n_f=4)}(\mu,m_b)\,,
\end{equation}
where $U^{(n_f=4)}(\mu,m_b)$ and $R$ are the $6\times 6$ submatrices from the corresponding quantities 
defined above. This time the vector $C(M_W)$ contains $C_1(M_W)$ to $C_6(M_W)$ where, as 
stated already, 
only two are non-vanishing. The solution of (\ref{eq::W}) can be found in~(C.16) 
in~\cite{Beneke:2001at}. The $1\times 6$ matrix $\kappa$ that describes the mixing into ${\cal O}_9$ is 
given by~\cite{Gambino:2003zm}
\begin{equation}
\kappa = \kappa^{(-1)} +\kappa^{(0)} a_s + \ldots\,,
\end{equation}
with
\begin{equation}
  \kappa^{(-1)T} = 
\left(
\begin{array}{c}
 -\frac{16}{9}q_1 \\
 -\frac{4}{3}q_1 \\
 -\frac{8}{3}q_2-8 q \\
 -\frac{32}{9}q_2 \\
 -\frac{128}{3}q_2-80 q \\
 -\frac{512}{9}q_2 \\
\end{array}
\right),
\end{equation}

\begin{equation}
  \kappa^{(0)T} = 
  \left(
\begin{array}{c}
 -\frac{136}{27} q_1-\frac{176}{243} q_2 \\
 \frac{128}{9} q_1+\frac{352}{81} q_2 \\
 \frac{4160}{81} q_2-32 q \\
 \left(-\frac{784}{243} +\frac{544}{81} n_f\right) q_2 +\frac{64}{3} q\\
 \frac{58112}{81} q_2-320 q \\
 \left(\frac{22784}{243} +\frac{4288}{81} n_f\right) q_2+\frac{608}{3} q\\
\end{array}
\right),
\end{equation}

\begin{eqnarray}
  \kappa^{(1)T} &=&
\left(
\begin{array}{c}
 -\frac{14999}{81} q_1-\left(\frac{72560}{6561} +\frac{1120}{2187} n_f \right)q_2
 +\frac{3152}{243} n_2 \bar q+\frac{7976}{243} q \\
 \frac{820}{27} q_1+\left(\frac{333688}{2187}+\frac{2240}{729} n_f\right) q_2-\frac{184}{81} n_2 \bar q 
+\frac{1832}{81} q \\
 \left(\frac{1524104}{2187}-\frac{44048}{729} n_f\right) q_2
 -\left(\frac{2636}{9}-\frac{176}{9} n_f\right) q \\
 \left(-\frac{1535926}{6561}+\frac{159620}{2187} n_f+\frac{608}{729} n_f^2\right) q_2
 +\left(\frac{1201}{27}  -\frac{32}{9} n_f \right)q \\
 \left(\frac{31433600}{2187}+\frac{15904}{729} n_f\right) q_2
 +\left(\frac{46552}{9} +\frac{2912}{9} n_f \right)q \\
 \left(-\frac{48510784}{6561}+\frac{3516560}{2187} n_f+\frac{15872}{729} n_f^2\right) q_2
 -\left(\frac{47624}{27}-\frac{1312}{27} n_f  \right) q \\
\end{array}
\right)\\
 &&+ \zeta (3)\left(
\begin{array}{c}
\frac{352}{9} q_1-\frac{640}{81} q_2\\
 -\frac{128}{3} q_1+\frac{1280}{27} q_2 \\
 \frac{256}{27} q_2+128 q \\
 \left(\frac{5056}{81}+\frac{1280}{27} n_f\right) q_2+\frac{160}{3} q \\
 \frac{4096}{27} q_2+1280 q\\
 \left(\frac{80896}{81} +\frac{12800}{27} n_f\right) q_2-\frac{512}{3} q\\
\end{array}
\right),
\end{eqnarray}
where $q=n_1q_1+n_2q_2$.
The initial condition for $C_9$ at the scale $m_b$ stems from the matching of the five-quark to 
the four-quark theory. The leading-order contribution arises from diagrams similar to the one in 
Fig.~\ref{diag::match}, but 
with the gluon exchanged by a photon and the quark-antiquark-pair by a lepton-pair. It is given 
by~\cite{Buras:1993dy}
\begin{eqnarray}
 C_9(m_b) = -\frac{8}{27}\left(C_1(m_b)+\frac{3}{4}C_2(m_b)\right).
\end{eqnarray}
The two-loop 
contributions consist solely of diagrams like the ones calculated in~\cite{Asatryan:2001zw}. Again, due 
to the expansion used there, we cannot use the results for our purpose and we will neglect these higher 
order contribution in our results.

We are now ready to present the results for the Wilson coefficients in 
Table~\ref{tab::Wilson}. It can be noted that the numerical results in the four-quark sector, $C_{1-6}$ 
at the scale $\mu=1.3$~GeV, are not much different than the ones for $b$-decays at the scale $\mu=m_b$. 
Only $C_1$ is about twice as large for charm decays, whereas $C_{2-6}$ are very similar. The main 
difference is observed for the coefficients $C_7^{\rm eff}$, $C_8^{\rm eff}$, $C_9$ and 
$C_{10}$. $C_7^{\rm eff}$ 
has a different sign and is roughly a factor of six smaller. $C_8^{\rm eff}$ is roughly a factor of
three smaller than in $b$-decays. Whereas $C_{10}$ is exactly zero to all orders in the strong 
coupling 
as explained above, also $C_9$ is an order of magnitude smaller for charm decays. Concerning the NNLL 
results, one has of course to bear in mind that we have neglected the two-loop matching conditions at 
the 
scale $\mu=m_b$.

One of us has already used the results presented here to perform a phenomenological 
analysis of $D$-decays~\cite{deBoer:2015boa}.

\begin{table}[ht!]
\centerline{\parbox{14cm}{\caption{\label{tab::Wilson}
Wilson coefficients at the scale $\mu=1.3\,$GeV in leading-log\-arithmic
(LL), next-to-leading-logarithmic (NLL) and next-to-next-to-leading-logarithmic 
(NNLL) order for $C_{1-6}$, $C_9$ and $C_{10}$. Input parameters 
are $\Lambda^{(4)}_{\overline{\rm MS}}=0.294$\,GeV,
$\Lambda^{(5)}_{\overline{\rm MS}}=0.214$\,GeV,
$\hat m_t(\hat m_t)=163.3$\,GeV, 
$M_W=80.4$\,GeV and $\hat m_b(\hat m_b)=4.18$\,GeV. 3-loop running is used 
for $\alpha_s$.}}}
\vspace{0.1cm}
\begin{center}
\begin{tabular}{|l|c|c|c|c|c|c|}
\hline\hline
\rule[-2mm]{0mm}{7mm}
 & $C_1$ & $C_2$ & $C_3$ & $C_4$ & $C_5$
 & $C_6$ \\
\hline
\rule[-0mm]{0mm}{4mm}
LL    & $-1.035$ & $1.094$ & $-0.004$ & $-0.061$ & $0.000$ & $0.001$ \\

NLL   & $-0.712$ & $1.038$ & $-0.006$ & $-0.093$ & $0.000$ & $0.001$ \\
NNLL   & $-0.633$ & $1.034$ & $-0.008$ & $-0.093$ & $0.000$ & $0.001$ \\
\hline
\rule[-2mm]{0mm}{7mm}
 & $C_7^{\rm eff}$ & $C_8^{\rm eff}$ & $C_9$ & $C_{10}$
 & $C_9^{\rm NNLL}$ &  $C_{10}^{\rm NNLL}$ \\
\hline
\rule[-0mm]{0mm}{4mm}
LL  & $0.078$ & $-0.055$ & $-0.098$ & 0
 & & \\
NLL & $0.051$ & $-0.062$ & $-0.309$ & 0
 & \raisebox{2.5mm}[-2.5mm]{$-0.488$} & \raisebox{2.5mm}[-2.5mm]{0} \\
\hline\hline
\end{tabular}
\end{center}
\end{table} 

\section{Effective Wilson coefficient $C_9^{\rm eff}$}

We will now, analogously to the case of $B$-physics, introduce the renormalisation-scheme independent 
effective ``Wilson coefficient'' $C_9^{\rm eff}$, which absorbs the universal 
long-distance effects from quark loops in perturbation theory~\cite{Misiak:1992bc}:
\begin{eqnarray}
 C_9^{\rm eff}(\mu,s) &=& (V^*_{cd} V_{ud}+V^*_{cs} V_{us})\left(C_9(\mu) + 
Y^{(ds)}(\mu,s)\right)\nonumber\\
 &&+ V^*_{cd} V_{ud}\, Y^{(d)}(\mu,s) + V^*_{cs} V_{us}\, Y^{(s)}(\mu,s)\,,
\end{eqnarray}
where $s=q^2$, $q=p-p'$, with the momentum $p$ and $p'$ of the incoming $c$- and outgoing $u$-quark, 
respectively. The functions $Y^{(i)}(\mu,s)$ are defined as
\begin{eqnarray}
 Y^{(d)}(\mu,s) &=& h(\mu,s,0) \left(\frac43 C_1(\mu)+C_2(\mu)\right),\\
 Y^{(s)}(\mu,s) &=& h(\mu,s,m_s) \left(\frac43 C_1(\mu)+C_2(\mu)\right),\\
 Y^{(ds)}(\mu,s) &=& -2 h(\mu,s,m_c) \left(7C_3(\mu)+\frac43 
C_4(\mu)+76C_5(\mu)+\frac{64}{3}C_6(\mu)\right)\nonumber\\
 && + h(\mu,s,m_s) \left(6C_3(\mu)+60C_5(\mu)\right)\nonumber\\
 && -\frac43 h(\mu,s,0) \left(6C_3(\mu)+2 C_4(\mu)+69C_5(\mu)+32C_6(\mu)\right)\nonumber\\
 &&+\frac{8}{3}\left(C_3(\mu)+10C_5(\mu)\right),
\end{eqnarray}
where
\begin{equation}
\label{eq:h}
h(\mu,s,m_q) = \frac{2}{9}\left(\ln\frac{m_q^2}{\mu^2} - \frac{2}{3}
- z \right)- \frac{1}{9} \,(2+z) \, B_0(s,m_q)\,,
\end{equation}
with $z=4m_q^2/s$, and
\begin{equation}
\label{eq:B0}
B_0(s,m_q) = - 2 \,\sqrt{\,|z-1|} \,
\left\{
\begin{array}{l}
\,\arctan\displaystyle{\frac{1}{\sqrt{z-1}}}
\qquad\quad z>1\\[0.4cm]
\,\ln\displaystyle{\frac{1+\sqrt{1-z}}{\sqrt{z}}} - \frac{i\pi}{2}
\quad z\leq 1
\end{array}
\right..
\end{equation}

We will not consider two-loop corrections to the matrix elements in this paper and concentrate on the 
one-loop corrections which have been dealt with in previous 
works~\cite{Fajfer:2001sa,Burdman:2001tf,Fajfer:2002gp,Paul:2011ar}. In all these papers a 
different operator basis was used. To compare those results with ours, one can simply use the formulae 
given in Appendix~\ref{app::a}.

In~\cite{Fajfer:2001sa,Burdman:2001tf} the findings of Inami and Lim~\cite{Inami:1980fz} were used to 
estimate the Wilson coefficient $C_9$ from electroweak theory without QCD. It was later pointed 
out by Fajfer et al.~\cite{Fajfer:2002gp} that this leads to a great overestimation of the decay width. 
We agree with the authors on that point. However, in 2011 Paul et al.~\cite{Paul:2011ar} argued that 
those results contain a sign error in the function analogous to our function $h$ defined in 
(\ref{eq:h}), which would invalidate the main arguments given in~\cite{Fajfer:2002gp}. We will therefore 
try to clarify this point again in a slightly different way than in~\cite{Fajfer:2002gp}.

Let us first look at the case of $B$-decays. To obtain the matching condition at the scale $M_W$ at 
leading order for the Wilson coefficient $C_9$, one first has to calculate penguin and box diagrams in 
full QCD. This calculation has been performed by Inami and Lim~\cite{Inami:1980fz}. The result contains 
logarithms of the form $\log (m_t/M_W)$ and $\log (m_c/M_W)$. The $u$-quark mass is set to zero and the 
corresponding IR-singularity is regularised dimensionally. Then the corresponding diagrams have to be 
computed within the effective theory. The Wilson coefficient has to be chosen in such a way that both 
calculations coincide at the scale $M_W$, where the matching can be performed at zero momentum transfer. 
The effective theory calculation thus leads to terms proportional to the $h$-function in (\ref{eq:h}) at 
$s=0$:
\begin{eqnarray}\label{eq::h0}
 h(\mu,0,m_q) = \frac{2}{9}\left(1+\log\frac{m_q^2}{\mu^2}\right).
\end{eqnarray}
Again, the $u$-quark mass is set to zero and therefore the corresponding diagram 
vanishes within dimensional regularisation. As the top-quark does not appear in the effective theory, 
the term containing $\log (m_t/M_W)$ can obviously not be reproduced unless it is contained in 
$C_9(M_W)$. The $\log (m_c/\mu)$ term in (\ref{eq::h0}) matches exactly the $\log (m_c/M_W)$ term from 
the full QCD calculation which leads to a $\log (\mu/M_W)$ term in $C_9(\mu\sim M_W)$, i.e. the explicit 
logarithms for the light quark masses in the full theory have {\em the same} sign as in the quark 
loop function $h(\mu,s,m_q)$. This is what is 
expected, as $m_c\ll M_W$, and the corresponding contributions are considered long-distance (as 
compared to the scale $M_W$) and should be reproduced within the effective theory and not be contained 
in the Wilson coefficient. In the actual matching calculation one of course sets $m_c$ to zero from the 
beginning which leads to the same result for $C_9(M_W)$.

In the case of $D$-decays, the roles of $t$-, $c$- and $u$-quarks are taken over by $b$-, $s$- and 
$d$-quarks. By the same reasoning as before, this time {\em all} the quark masses can be set to zero in 
the matching calculation which immediately leads to vanishing $C_9(M_W)$ due to the unitarity of the 
CKM-matrix. When Paul et al.~\cite{Paul:2011ar} state that the logarithms in the Inami-Lim term and in 
the effective QCD corrections have to have a different sign, it should be clear from the above 
considerations that this cannot be true. Moreover, the function $h$ has a smooth limit for $m_q\to 0$ at 
$s\ne 0$:
\begin{eqnarray}
 h(\mu,s,0) = -\frac{2}{27}\left(2+3\pi i-3\log\frac{s}{\mu^2}\right).
\end{eqnarray}
If the logarithm in the Inami-Lim term were to cancel the explicit logarithm in the first term in 
(\ref{eq:h}), the whole contribution would contain a logarithmic divergence for vanishing quark masses 
at $s\ne 0$.

\section{Conclusions and outlook}

In this paper we have presented the calculation of Wilson coefficients for the weak effective 
Hamiltonian relevant for rare semileptonic decays of $D$-mesons at NNLL order which is required to 
perform 
an analysis of those decays at first order in the strong coupling $\alpha_s$. The calculation is very 
similar to the analogous one for $B$-meson decays. 
The main difference arises through the necessity to perform a two-step matching, as one has to cross the 
$b$-quark threshold while evolving the renormalisation scale from the high scale $M_W$ down to the 
charm-mass 
scale. The corresponding anomalous dimensions and initial conditions at $M_W$ could be taken from the 
results known in the $B$-meson sector, with the obvious replacements of quark charges and number of 
flavours within the effective theory. We tried to clarify some misunderstanding present in the 
literature concerning the correct matching at the scale $M_W$.

As mentioned in the introduction, due to the specific CKM and mass structure of charm FCNCs, the 
short distance contributions within the SM can differ by several orders of magnitude depending on which 
corrections are taken into account. We have seen that many of the Wilson coefficients are very similar 
to the ones for $b$-decays. Only $C_9$ differs by one order of magnitude and $C_{10}$ 
is zero. To fully exploit the SM short-distance contributions one of course has to take into 
account the hadronic matrix elements within the effective theory. This will be done at the same order in 
the strong coupling $\alpha_s$ in a 
future publication~\cite{FMS}.

\section*{Acknowledgements}

We thank Thorsten Feldmann for discussions and careful reading of the manuscript and Martin Gorbahn 
for providing us with the full parameter dependence of the anomalous dimension matrices. This work is 
supported in parts by the Bundesministerium f\"ur Bildung und Forschung (BMBF), and the
Deutsche Forschungsgemeinschaft (DFG) within research unit FOR 1873 (``QFET'').

\begin{appendix}

\section{Alternative operator bases}
\label{app::a}

\begin{table}[t]
\centerline{\parbox{14cm}{\caption{\label{tab::Wilson2}
``Barred'' Wilson coefficients $\bar C_{1-6}$ at the scale $\mu=1.3\,$GeV in leading-logarithmic
(LL), next-to-leading-logarithmic (NLL) and next-to-next-to-leading-logarithmic 
(NNLL) order. Input parameters are the same as in Tab.~\ref{tab::Wilson}.}}}
\vspace{0.1cm}
\begin{center}
\begin{tabular}{|l|c|c|c|c|c|c|}
\hline\hline
\rule[-2mm]{0mm}{7mm}
 & $\bar{C}_1$ & $\bar{C}_2$ & $\bar{C}_3$ & $\bar{C}_4$ & $\bar{C}_5$
 & $\bar{C}_6$ \\
\hline
\rule[-0mm]{0mm}{4mm}
LL    & $-0.517$ & $1.266$ & $0.010$ & $-0.025$ & $0.007$ & $-0.029$ \\

NLL   & $-0.356$ & $1.157$ & $0.014$ & $-0.042$ & $0.010$ & $-0.045$ \\
NNLL   & $-0.317$ & $1.140$ & $0.013$ & $-0.040$ & $0.009$ & $-0.045$ \\
\hline\hline
\end{tabular}
\end{center}
\end{table}

For comparison with previous work we will introduce ``barred'' coefficients 
$\bar{C}_i$ (for  $i=1,\ldots,6$), defined by the following linear combinations of the Wilson 
coefficients $C_i$~\cite{Beneke:2001at}:
\begin{eqnarray}
\bar{C}_1 &=& \frac{1}{2} \,C_1\,,\nonumber\\
\bar{C}_2 &=& C_2-\frac{1}{6}\,C_1\,,\nonumber\\
\bar{C}_3 &=& C_3-\frac{1}{6}\,C_4+16\,C_5-\frac{8}{3}\,C_6\,,\nonumber\\
\bar{C}_4 &=& \frac{1}{2}\,C_4+8\,C_6\,,\nonumber\\
\bar{C}_5 &=& C_3-\frac{1}{6}\,C_4+4\,C_5-\frac{2}{3}\,C_6\,,\nonumber\\
\bar{C}_6 &=& \frac{1}{2}\,C_4+2\,C_6\,.
\end{eqnarray}
The linear combinations are chosen such that the $\bar{C}_i$ coincide {\em at leading logarithmic order} 
with the Wilson coefficients in the standard basis \cite{Buchalla:1995vs}. Numerical values for the 
coefficients are 
listed in Tab.~\ref{tab::Wilson2}. These definitions hold to all orders in perturbation theory. The 
``barred'' coefficients are related to those defined in~cite{Buchalla:1995vs} by~\cite{Chetyrkin:1997gb}
\begin{equation}
\bar{C}_i = C^{\rm BBL}_i +\frac{\alpha_s}{4\pi}\,T_{ij}\, C^{\rm
  BBL}_j + O(\alpha_s^2)\,,
\end{equation}
where 
\begin{equation}
T = \left(\begin{array}{cccccc}
\frac{7}{3} & 2 & 0 & 0 & 0 & 0 \\[0.1cm]
1 & -\frac{2}{3} & 0 & 0 & 0 & 0 \\[0.1cm]
0 & 0 & -\frac{178}{27} & -\frac{4}{9} & \frac{160}{27} & \frac{13}{9}
\\[0.1cm]
0 & 0 & \frac{34}{9} & \frac{20}{3} & -\frac{16}{9} & -\frac{13}{3}
\\[0.1cm]
0 & 0 & \frac{164}{27} & \frac{23}{9} & -\frac{146}{27} & -\frac{32}{9}
\\[0.1cm]
0 & 0 & -\frac{20}{9} & -\frac{23}{3} & \frac{2}{9} & \frac{16}{3}
\end{array}\right).
\end{equation}

\end{appendix}

\end{document}